\documentclass[fleqn,twoside]{article}
\usepackage{espcrc2}

\usepackage{graphicx}

\usepackage{cite}

\def\NPPS{Nucl. Phys. B (Proc. Suppl.)}

\def\MP{Int. J. Mod. Phys.}

\newcommand{\eqn}[1]{(\ref{#1})}
\newcommand{\be}{\begin{equation}}
\newcommand{\ee}{\end{equation}}
\newcommand{\no}{\nonumber}
\newcommand{\bel}[1]{\be\label{#1}}
\newcommand{\ba}{\begin{array}{c}}
\newcommand{\bat}{\begin{array}{cc}}
\newcommand{\ea}{\end{array}}
\newcommand{\beqn}{\begin{eqnarray}}
\newcommand{\eeqn}{\end{eqnarray}}

\newcommand{\bi}{\begin{itemize}}
\newcommand{\ei}{\end{itemize}}

\newcommand{\cO}{{\cal O}}

\newcommand{\cJ}{{\cal J}}

\newcommand{\AmS}{{\protect\the\textfont2
  A\kern-.1667em\lower.5ex\hbox{M}\kern-.125emS}}

\title{QCD Description of Hadronic Tau Decays}

\author{A. Pich\address{Departament de F\'{\i}sica Te\`orica, IFIC,
Univ. Val\`encia--CSIC, Apt. 22085, E-46071 Val\`encia, Spain}
}

\begin{document}

\begin{abstract}
The QCD analysis of hadronic $\tau$ decays is reviewed and a summary of the present phenomenological status is presented.
The following topics are discussed:
the determination of $\alpha_s(m_\tau^2) = 0.338 \pm 0.012$ from the inclusive $\tau$ hadronic width,
the measurement of $|V_{us}|$ through the Cabibbo-suppressed decays
of the $\tau$, and the extraction of chiral-perturbation-theory couplings from the spectral tau data.
\vspace{1pc}
\end{abstract}

\maketitle

\section{THEORETICAL FRAMEWORK}

The hadronic $\tau$ decays turn out to be a beautiful
laboratory for studying strong interaction effects at low energies \cite{Kazimierz,Reviews,taurev98}.
The $\tau$ is the only known lepton massive enough to decay into hadrons.
Its semileptonic decays are then ideally suited to investigate the hadronic weak currents.

The inclusive character of the total $\tau$ hadronic width renders
possible an accurate calculation of the ratio
\cite{BR:88,NP:88,BNP:92,LDP:92a,QCD:94}
$$
 R_\tau \equiv { \Gamma [\tau^- \to \nu_\tau
 \,\mathrm{hadrons}] \over \Gamma [\tau^- \to \nu_\tau e^-
 {\bar \nu}_e] } \, =\,
 R_{\tau,V} + R_{\tau,A} + R_{\tau,S}\, .
$$ 
The theoretical analysis involves the two-point correlation functions for
the vector $\, V^{\mu}_{ij} = \bar{\psi}_j \gamma^{\mu} \psi_i \, $
and axial-vector
$\, A^{\mu}_{ij} = \bar{\psi}_j \gamma^{\mu} \gamma_5 \psi_i \,$
colour-singlet quark currents ($i,j=u,d,s$):
\be\label{eq:pi_v}
\Pi^{\mu \nu}_{ij,\cJ}(q) \equiv
 i \int d^4x \, e^{iqx}
\langle 0|T(\cJ^{\mu}_{ij}(x) \cJ^{\nu}_{ij}(0)^\dagger)|0\rangle  ,
\ee
which have the Lorentz decompositions
\beqn\label{eq:lorentz}
\Pi^{\mu \nu}_{ij,\cJ}(q) & \!\!\! = & \!\!\!
  \left( -g^{\mu\nu} q^2 + q^{\mu} q^{\nu}\right) \: \Pi_{ij,\cJ}^{(1)}(q^2)
\no\\
  &\!\!\! & \!\!\! +\;   q^{\mu} q^{\nu} \, \Pi_{ij,\cJ}^{(0)}(q^2) \, ,
\eeqn
where the superscript $(J=0,1)$ denotes the angular momentum in the hadronic rest frame.

The imaginary parts of 
$\, \Pi^{(J)}_{ij,\cJ}(q^2) \, $
are proportional to the spectral functions for hadrons with the corresponding
quantum numbers.  The hadronic decay rate of the $\tau$
can be written as an integral of these spectral functions
over the invariant mass $s$ of the final-state hadrons:
\beqn\label{eq:spectral}
R_\tau  &\!\!\! = &\!\!\!
12 \pi \int^{m_\tau^2}_0 {ds \over m_\tau^2 } \,
 \left(1-{s \over m_\tau^2}\right)^2
\no\\ &\!\!\! \times &\!\!\!
\biggl[ \left(1 + 2 {s \over m_\tau^2}\right)
 \mbox{\rm Im} \Pi^{(1)}(s)
 + \mbox{\rm Im} \Pi^{(0)}(s) \biggr]  .
\eeqn
The appropriate combinations of correlators are
\beqn\label{eq:pi}
\Pi^{(J)}(s)  &\!\!\! \equiv  &\!\!\!
  |V_{ud}|^2 \, \left( \Pi^{(J)}_{ud,V}(s) + \Pi^{(J)}_{ud,A}(s) \right)
\no\\ &\!\!\! + &\!\!\!
|V_{us}|^2 \, \left( \Pi^{(J)}_{us,V}(s) + \Pi^{(J)}_{us,A}(s) \right).
\eeqn
 The contributions coming from the first two terms correspond to
$R_{\tau,V}$ and $R_{\tau,A}$ respectively, while
$R_{\tau,S}$ contains the remaining Cabibbo-suppressed contributions.

The integrand in Eq.~(\ref{eq:spectral}) cannot be calculated at present from QCD.
Nevertheless the integral itself can be calculated systematically by exploiting
the analytic properties of the correlators $\Pi^{(J)}(s)$. They are analytic
functions of $s$ except along the positive real $s$-axis, where their
imaginary parts have discontinuities.
$R_\tau$ can then be written as a contour integral
in the complex $s$-plane running
counter-clockwise around the circle $|s|=m_\tau^2$ \cite{BNP:92}:
\beqn\label{eq:circle}
 R_\tau &\!\!\!\! =&\!\!\!
6 \pi i \oint_{|s|=m_\tau^2} {ds \over m_\tau^2} \,
 \left(1 - {s \over m_\tau^2}\right)^2
\no \\ &\!\!\!\!\times &\!\!\!\!
\left[ \left(1 + 2 {s \over m_\tau^2}\right) \Pi^{(0+1)}(s)
         - 2 {s \over m_\tau^2} \Pi^{(0)}(s) \right] \! .
\eeqn
This expression requires the correlators only for
complex $s$ of order $m_\tau^2$, which is significantly larger than the scale
associated with non-perturbative effects.
Using the Operator Product Expansion (OPE),
$\Pi^{(J)}(s) = \sum_{D} C_D^{(J)}/ (-s)^{D/2}$,
to evaluate the contour integral, $R_\tau$
can be expressed as an expansion in powers of $1/m_\tau^2$.
%
The uncertainties associated with the use of the OPE near the
time-like axis are heavily suppressed by the presence in (\ref{eq:circle})
of a double zero at $s=m_\tau^2$.

In the chiral limit ($m_{u,d,s} = 0$),
the vector and axial-vector currents are conserved.
This implies  $s \,\Pi^{(0)}(s) = 0$. Therefore, only the correlator
$\Pi^{(0+1)}(s)$ contributes to Eq.~(\ref{eq:circle}).
Since $(1 - x)^2 (1 + 2 x) =
1 - 3 x^2 + 2x^3$ [$x\equiv s/m_\tau^2$],
Cauchy's theorem guarantees that, up to tiny logarithmic running corrections, the only non-perturbative
contributions to the circle integration in (\ref{eq:circle}) originate from operators of
dimensions $D=6$ and 8. The usually leading $D=4$ operators can only contribute to $R_\tau$ with an
additional suppression factor of $\cO(\alpha_s^2)$, which makes their effect negligible \cite{BNP:92}.

\section{DETERMINATION OF\ {\large $\mathbf{\alpha_s}$}}

\begin{table*}[tb]\centering
\caption{$\cO(\alpha_s^4)$ determinations of $\alpha_s(m_\tau^2)$. The assumed value of $\delta_{\mathrm{P}}$ is also given
(if quoted by the authors).}
\label{tab:alpha-pub-values}
\newcommand{\cc}[1]{\multicolumn{1}{c}{#1}}
\renewcommand{\tabcolsep}{1.pc} 
\renewcommand{\arraystretch}{1.2} 
\begin{tabular}{@{}lllll}
\hline
\cc{Reference} & \cc{Method} & \cc{$\delta_{\mathrm{P}}$} & \cc{$\alpha_s(m_\tau^2)$}  & \cc{$\alpha_s(M_Z^2)$}
\\ \hline
Baikov et al. 
\cite{BChK:08} & {\small CIPT, FOPT} & $0.1998 \pm 0.0043$ & $0.332\pm 0.016$ & $0.1202\pm 0.0019$
\\
Davier et al. \cite{DDMHZ:08} & {\small CIPT} & $0.2066\pm 0.0070$ & $0.344\pm 0.009$ & $0.1212\pm 0.0011$
\\
Beneke-Jamin \cite{BJ:08} & {\small BSR + FOPT} & $0.2042\pm 0.0050$ & $0.316\pm 0.006$ & $0.1180\pm 0.0008$
\\
Maltman-Yavin \cite{MY:08} & {\small PWM + CIPT} & \cc{---} & $0.321 \pm 0.013$ & $0.1187\pm 0.0016$
\\
Menke \cite{ME:09} & {\small CIPT, FOPT} 
& $0.2042\pm 0.0050$ & $0.342\: {}^{+\: 0.011}_{-\: 0.010}$  & $0.1213\pm 0.0012$
\\
Caprini-Fischer \cite{CF:09} & {\small BSR + CIPT} & $0.2042\pm 0.0050$ & $0.320\: {}^{+\: 0.011}_{-\: 0.009}$ & \cc{---}
\\
Cveti\v{c} et al. \cite{CLMV:10} & $\beta_{\mathrm{exp}}$ + CIPT &  $0.2040\pm 0.0040$ & $0.341\pm 0.008$ & $0.1211\pm 0.0010$
\\
Pich \cite{Kazimierz} & {\small CIPT} & $0.2038\pm 0.0040$ & $0.342\pm 0.012$ & $0.1213\pm 0.0014$
\\
\hline
\end{tabular}
\end{table*}

The Cabibbo-allowed combination $R_{\tau,V+A}$
can be written as \cite{BNP:92}
\begin{equation}\label{eq:Rv+a}
 R_{\tau,V+A} \, =\, N_C\, |V_{ud}|^2\, S_{\mathrm{EW}} \left\{ 1 +
 \delta_{\mathrm{P}} + \delta_{\mathrm{NP}} \right\} ,
\end{equation}
where $N_C=3$ is the number of quark colours
and $S_{\mathrm{EW}}=1.0201\pm 0.0003$ contains the
electroweak radiative corrections \cite{MS:88,BL:90,ER:02}.
The dominant correction ($\sim 20\%$) is the perturbative QCD
contribution $\delta_{\mathrm{P}}$, which is already known to
$O(\alpha_s^4)$ \cite{BNP:92,BChK:08}.
%
Quark mass effects  \cite{BNP:92,PP:99,BChK:05}
are tiny for the Cabibbo-allowed current
and amount to a negligible correction smaller than $10^{-4}$ \cite{BNP:92,BJ:08}.

Non-perturbative contributions are suppressed by six powers of the
$\tau$ mass \cite{BNP:92} and, therefore, are very small. Their
numerical size has been determined from the invariant-mass
distribution of the final hadrons in $\tau$ decay, through the study
of weighted integrals \cite{LDP:92b},
\begin{equation}\label{eq:moments}
 R_{\tau}^{kl} \,\equiv\, \int_0^{m_\tau^2} ds\, \left(1 - {s\over
 m_\tau^2}\right)^k\, \left({s\over m_\tau^2}\right)^l\, {d
 R_{\tau}\over ds} \, ,
\end{equation}
which can be calculated theoretically in the same way as $R_{\tau}$,
but are more sensitive to OPE corrections.
The predicted suppression \cite{BNP:92} of the non-perturbative
corrections to $R_\tau$ has been confirmed by ALEPH \cite{ALEPH:05}, CLEO
\cite{CLEO:95} and OPAL \cite{OPAL:98}. The most recent analysis gives
\cite{DHZ:05}
\begin{equation}\label{eq:del_np}
 \delta_{\mathrm{NP}} \, =\, -0.0059\pm 0.0014 \, .
\end{equation}

The QCD prediction for $R_{\tau,V+A}$ is then completely dominated
by $\delta_{\mathrm{P}}$; 
non-perturbative effects being
smaller than the perturbative uncertainties from uncalculated
higher-order corrections.
Assuming lepton universality, the measured values of the $\tau$ lifetime and leptonic branching ratios imply
$R_\tau = 3.6291\pm 0.0086$ \cite{HFAG}. Subtracting the Cabibbo-suppressed
contribution $R_{\tau,S}= 0.1613 \pm 0.0028$ \cite{HFAG}, one obtains
$R_{\tau,V+A} = 3.4678\pm 0.0090$. Using $|V_{ud}| = 0.97425\pm 0.00022$ \cite{PDG}
and \eqn{eq:del_np}, the pure perturbative contribution to $R_\tau$ is determined to be:
\bel{eq:delta_P}
\delta_{\mathrm{P}} = 0.1997 \pm 0.0035 \, .
\ee

The predicted value of $\delta_{\mathrm{P}}$ turns out to be very sensitive
to $\alpha_s(m_\tau^2)$, allowing for an accurate
determination of the fundamental QCD coupling \cite{NP:88,BNP:92}.
The calculation of the $\cO(\alpha_s^4)$ contribution \cite{BChK:08}
has triggered a renewed theoretical interest on the $\alpha_s(m_\tau^2)$ determination, since it allows to
push the accuracy to the four-loop level.
However, as shown in Table~\ref{tab:alpha-pub-values}, the recent theoretical analyses slightly disagree on the final result.
The differences are larger than the claimed $\cO(\alpha_s^4)$ accuracy and
originate in the different inputs or theoretical procedures which have been adopted.

\subsection{Perturbative contribution to $\boldmath R_\tau$}

%
\begin{table*}[tb]\centering
\caption{ Exact results
for $A^{(n)}(\alpha_s)$ ($n\le 4$) at different $\beta$-function approximations,
and corresponding values of \ $\delta_{\mathrm{P}} = \sum_{n=1}^4\, K_n\, A^{(n)}(\alpha_s)$,
for $a_\tau\equiv\alpha_s(m_\tau^2)/\pi=0.11$. The last row shows the FOPT estimates at $\cO(a_\tau^4)$.}
\label{tab:Afun}
\newcommand{\cc}[1]{\multicolumn{1}{c}{#1}}
\renewcommand{\tabcolsep}{1.5pc} 
\renewcommand{\arraystretch}{1.2} 
\begin{tabular}{@{}llllll}
\hline
& \cc{$A^{(1)}(\alpha_s)$} & \cc{$A^{(2)}(\alpha_s)$} & \cc{$A^{(3)}(\alpha_s)$}
 & \cc{$A^{(4)}(\alpha_s)$} & 
 \cc{$\delta_{\mathrm{P}}$}
\\ \hline
$\beta_{n>1}=0$ & $0.14828$ & $0.01925$ & $0.00225$ & $0.00024$ & 
$0.20578$ \\
$\beta_{n>2}=0$ & $0.15103$ & $0.01905$ & $0.00209$ & $0.00020$ & 
$0.20537$ \\
$\beta_{n>3}=0$ & $0.15093$ & $0.01882$ & $0.00202$ & $0.00019$ & 
$0.20389$ \\
$\beta_{n>4}=0$ & $0.15058$ & $0.01865$ & $0.00198$ & $0.00018$ & 
$0.20273$
\\ \hline
$\cO(a_\tau^4)$ &  $0.16115$ & $0.02431$ & $0.00290$ & $0.00015$ & 
$0.22665$
\\ \hline
\end{tabular}
\end{table*}
%

In the chiral limit, the result is more conveniently expressed in terms of the
logarithmic derivative of the two-point correlation function of the vector (axial) current,
$\Pi(s)=\frac{1}{2}\,\Pi^{(0+1)}(s)$,
which satisfies an homogeneous renormalization--group equation:
\be\label{eq:d}
 D(s)  \equiv
- s {d \over ds } \Pi(s)
=  {1\over 4 \pi^2} \sum_{n=0}  K_n
\left( {\alpha_s(-s)\over \pi}\right)^n \! .
\ee
With the choice of renormalization scale $\mu^2= - s$ all logarithmic corrections,
proportional to powers of $\log{(-s/\mu^2)}$, have been summed into the running coupling.
For three flavours, the known coefficients take the values:
$K_0 = K_1 = 1$; $K_2 = 1.63982$; 
$K_3(\overline{MS}) = 6.37101$ and $K_4(\overline{MS}) =49.07570$  \cite{BChK:08}.

The perturbative component of $R_\tau$ is given by
\be\label{eq:r_k_exp}
\delta_{\mathrm{P}} \, =\,
\sum_{n=1}  K_n \, A^{(n)}(\alpha_s) \, ,
\ee
where the functions \cite{LDP:92a}
\beqn\label{eq:a_xi}
A^{(n)}(\alpha_s) &\!\!\! = &\!\!\! {1\over 2 \pi i}\,
\oint_{|s| = m_\tau^2} {ds \over s} \,
  \left({\alpha_s(-s)\over\pi}\right)^n
\no\\ &\!\!\!\times  &\!\!\!
 \left( 1 - 2 {s \over m_\tau^2} + 2 {s^3 \over m_\tau^6}
         - {s^4 \over  m_\tau^8} \right)
\eeqn
are contour integrals in the complex plane, which only depend on
$a_\tau\equiv\alpha_s(m_\tau^2)/\pi$. Using the exact solution
(up to unknown $\beta_{n>4}$ contributions) for $\alpha_s(-s)$
given by the renormalization-group $\beta$-function equation,
they can be numerically computed with a very high accuracy \cite{LDP:92a}.
Table~\ref{tab:Afun} gives the numerical values
for $A^{(n)}(\alpha_s)$ ($n\le 4$) obtained at the one-, two-, three- and four-loop
approximations (i.e. $\beta_{n>1}=0$, $\beta_{n>2}=0$, $\beta_{n>3}=0$ and $\beta_{n>4}=0$,
respectively), together with the corresponding results for $\delta_{\mathrm{P}} = \sum_{n=1}^4\, K_n\, A^{(n)}(\alpha_s)$,
taking $a_\tau=0.11$.
The perturbative convergence is very good and the results are stable under changes of the renormalization scale.
The error induced by the truncation of the $\beta$ function at fourth order can be conservatively estimated
through the variation of the results at five loops, assuming $\beta_5 =\pm \beta_4^2/\beta_3 = \mp 443$,
i.e. a geometric growth of the $\beta$ function.

Higher-order contributions to the Adler function $D(s)$ will be taken into account
adding the fifth-order term $ K_5\, A^{(5)}(\alpha_s)$ with $K_5 = 275\pm 400$. Moreover, we will include
the 5-loop variation with changes of the renormalization scale in the range $\mu^2/(-s) \in [0.5,1.5]$.
Adopting this very conservative procedure, the experimental value of $\delta_{\mathrm{P}}$ given in Eq.~\eqn{eq:delta_P}
implies
\bel{eq:alpha-result}
\alpha_s(m_\tau^2) = 0.338 \pm 0.012\, .
\ee
The result is slightly lower than the one given in Ref.~\cite{Kazimierz}, due to the smaller value of $\delta_{\mathrm{P}}$.

The strong coupling measured at the $\tau$ mass scale is
significantly larger than the values obtained at higher energies.
From the hadronic decays of the $Z$, one gets $\alpha_s(M_Z^2) =
0.1190\pm 0.0027$ \cite{LEPEWWG}, which differs from $\alpha_s(m_\tau^2)$
by $18\,\sigma$. 
After evolution up to the scale $M_Z$ \cite{Rodrigo:1998zd}, the strong
coupling constant in (\ref{eq:alpha-result}) decreases to
\begin{equation}\label{eq:alpha_z}
 \alpha_s(M_Z^2)  \, =\,  0.1209\pm 0.0014 \, ,
\end{equation}
in excellent agreement with the direct measurements at the $Z$ peak
and with a better accuracy. The comparison of these two
determinations of $\alpha_s$ in two very different energy regimes, $m_\tau$
and $M_Z$, provides a beautiful test of the predicted running of the
QCD coupling; i.e., a very significant experimental verification of
{\it asymptotic freedom}.

\subsection{Fixed-order perturbation theory}

The integrals $A^{(n)}(\alpha_s)$ can be expanded in powers of $a_\tau$, 
$A^{(n)}(\alpha_s) = a_\tau^n + \cO(a_\tau^{n+1})$. One recovers in this way the naive perturbative expansion \cite{LDP:92a}
\bel{eq:d_fopt}
\delta_{\mathrm{P}}\, =\, \sum_{n=1}\,  (K_n + g_n) \, a_\tau^n \,\equiv\,
\sum_{n=1}\,  r_n \, a_\tau^n \, .
\ee
This approximation is known as {\it fixed-order perturbation theory} (FOPT), while
the improved expression \eqn{eq:r_k_exp}, keeping the non-expanded values of $A^{(n)}(\alpha_s)$,
is usually called {\it contour-improved perturbation theory} (CIPT) \cite{LDP:92a,PI:92}.

As shown in the last row of Table~\ref{tab:Afun}, even at $\cO(a_\tau^4)$, FOPT gives a rather bad approximation to the
integrals $A^{(n)}(\alpha_s)$, overestimating $\delta_{\mathrm{P}}$ by 12\% at $a_\tau = 0.11$.
The long running of $\alpha_s(-s)$ along the circle $|s|=m_\tau^2$ generates very large $g_n$ coefficients,
which depend on $K_{m<n}$ and $\beta_{m<n}$ \cite{LDP:92a}:
$g_1=0$, $g_2 =  3.56$, $g_3 = 19.99$, $g_4 = 78.00$, $g_5 = 307.78$. These corrections are much larger than the
original $K_n$ contributions and lead to values of $\alpha_s(m_\tau^2)$ smaller than \eqn{eq:alpha-result}.
FOPT suffers from a large renormalization-scale dependence \cite{LDP:92a} and its
actual uncertainties are much larger than usually estimated \cite{ME:09}.

The origin of this bad behaviour can be understood analytically at one loop \cite{LDP:92a}. In FOPT one makes within
the contour integral the series expansion ($\log{(-s/m_\tau^2)} = i\phi$, $\phi\in [-\pi, \pi]$)
\bel{eq:fopt_ap}
\frac{\alpha_s(-s)}{\pi} \approx \frac{a_\tau}{1-i\beta_1 a_\tau\phi/2}\approx a_\tau \sum_n \left(
\frac{i}{2}\beta_1 a_\tau\phi\right)^n \! ,
\ee
which is only convergent for $a_\tau < 0.14$. At the four-loop level the radius of convergence is slightly smaller than the physical value of $a_\tau$. Thus, FOPT gives rise to a pathological non-convergent series. The long running along the circle makes compulsory to resum the large logarithms, $\log^n{(-s/m_\tau^2)}$, using the renormalization group. This is precisely what CIPT does.

\subsection{Renormalon hypothesis}

The perturbative expansion of the Adler function is expected to be an asymptotic series. If its Borel transform,
$B(t)\equiv\sum_{n=0} K_{n+1} t^n/n!$, were well-behaved, one could define $D(s)$ through the Borel integral
\bel{eq:borel}
D(s) = \frac{1}{4\pi^2}\left\{ 1 + \int_0^\infty dt\, \mathrm{e}^{-\pi t/\alpha_s(s)} B(t)\right\} .
\ee
However, $B(t)$ has pole singularities at positive (infrared renormalons) and negative (ultraviolet renormalons) integer values of the variable
$u\equiv -\beta_1 t/2$, with the exception of $u=1$ \cite{BE:99}.
The infrared renormalons at $u=+n$ are related to OPE corrections of dimension $D=2n$.
The renormalon poles closer to the origin dominate the large-order behaviour of $D(s)$.

It has been argued that, once in the asymptotic regime (large $n$), the renormalonic behaviour of the $K_n$ coefficients could induce
cancelations with the running $g_n$ corrections, which would be missed by
CIPT. In that case, FOPT could approach faster the `true' result provided by the Borel summation
of the full renormalon series (BSR) \cite{BJ:08}.
This happens actually in the large--$\beta_1$ limit \cite{BBB:95,NE:96},
which however does not approximate well the known $K_n$ coefficients.
A model of higher-order corrections with this behaviour has been recently advocated \cite{BJ:08}.
The model mixes three different types of renormalons ($n=-1$, 2 and 3) plus a linear polynomial.
It contains 5 free parameters which are determined by the known values of $K_{1,2,3,4}$ and
the assumption $K_5=283$. One gets in this way a larger $\delta_{\mathrm{P}}$, implying
a smaller value for $\alpha_s(m_\tau^2)$. The result looks however model dependent \cite{DM:10}.

The implications of a renormalonic behaviour have been put on more solid grounds,
using an optimal conformal mapping in the Borel plane, which achieves the best asymptotic rate of convergence,
and properly implementing the CIPT procedure within the Borel transform \cite{CF:09}. Assuming that the known fourth-order series is already governed by the $u=-1$ and $u=2$ renormalons, the conformal mapping generates a full series expansion ($K_5=256$, $K_6=2929$ \ldots) which results, after Borel summation, in a larger value of $\delta_{\mathrm{P}}$; i.e. the $K_{n>4}$ terms give a positive contribution to $\delta_{\mathrm{P}}$ implying a smaller $\alpha_s(m_\tau^2)$ \cite{CF:09}.

Renormalons provide an interesting guide to possible higher-order corrections,
making apparent that the associated uncertainties have to be carefully estimated. However, one should keep in mind
the adopted assumptions. In fact, there are no visible signs of renormalonic behaviour in the presently known series: the $n=-1$ ultraviolet renormalon is expected to dominate the asymptotic regime, implying an alternating series,
while all known $K_n$ coefficients have the same sign. One could either assume that renormalons only become relevant at higher
orders, for instance at $n=7$, and apply the conformal mapping with arbitrary input values for $K_5$ and $K_6$. Different assumptions about these two unknown coefficients would result in different central values for $\alpha_s(m_\tau^2)$.

A different reshuffling of the perturbative series, not related to renormalons, has been recently proposed \cite{CLMV:10}.
Instead of the usual expansion in powers of the strong coupling, one expands in terms of the $\beta$ function and its derivatives
($\beta_{\mathrm{exp}}$), which effectively results in a different estimate of higher-order corrections. One gets in this way a weaker dependence on the renormalization scale and a value of $\alpha_s(m_\tau^2)$ similar to the standard CIPT result.

\subsection{Non-perturbative corrections}

At the presently achieved precision, one should worry about the small non-perturbative corrections. In fact, a proper definition of the infrared renormalon contributions is linked to the corresponding OPE corrections with $D=2n$.
A recent re-analysis of the ALEPH data \cite{MY:08}, with pinched-weight moments of the hadronic distribution (PWM) and CIPT,
obtains $\alpha_s(m_\tau^2) = 0.321\pm 0.013$. This smaller value originates in a different estimate of the
non-perturbative contributions. Unfortunately, Ref.~\cite{MY:08} does not quote any explicit values for $\delta_{\mathrm{NP}}$ and
$\delta_{\mathrm{P}}$. From the information given in that reference, I deduce
$\delta_{\mathrm{NP}}=0.012 \pm 0.018$. Although compatible with (\ref{eq:del_np}), the central value is larger and has the opposite sign. This shift implies a smaller $\delta_{\mathrm{P}}$ and, therefore, a slightly smaller strong coupling.

The so-called duality violation effects, i.e. the uncertainties associated with
the use of the OPE to approximate the exact correlator, have been also investigated
\cite{DDMHZ:08,CGP:09}. Owing to the presence in (\ref{eq:circle})
of a double zero at $s=m_\tau^2$, these effects are quite suppressed in $R_\tau$.
They are smaller than the errors induced by $\delta_{\mathrm{NP}}$, which are in turn subdominant with
respect to the leading perturbative uncertainties.

\section{$\mathbf{|V_{us}|}$ DETERMINATION}

The separate measurement of the $|\Delta S|=0$ and $|\Delta S|=1$
tau decay widths provides a very clean determination of $V_{us}\,$
\cite{GJPPS:05,PI:07b}.
To a first approximation the Cabibbo mixing can be directly obtained
from experimental measurements, without any theoretical input.
Neglecting the small SU(3)-breaking corrections from the $m_s-m_d$
quark-mass difference, one gets:
$$ 
 |V_{us}|^{\mathrm{SU(3)}} =\: |V_{ud}| \left(\frac{R_{\tau,S}}{R_{\tau,V+A}}\right)^{1/2}
 =\: 0.210\pm 0.002\, .
$$ 
The new branching ratios measured by BaBar and Belle are all smaller than the previous
world averages, which translates into a smaller value of $R_{\tau,S}$ and $|V_{us}|$.
For comparison, the previous value $R_{\tau,S}=0.1686\pm 0.0047$ \cite{DHZ:05} resulted in $|V_{us}|^{\mathrm{SU(3)}}=0.215\pm 0.003$.

This rather remarkable determination is only slightly shifted by
the small SU(3)-breaking contributions induced by the strange quark mass.
These effects can be 
estimated through a QCD analysis of the differences
\cite{GJPPS:05,PI:07b,PP:99,ChDGHPP:01,ChKP:98,KKP:01,MW:06,KM:00,MA:98,BChK:05}
\begin{equation}
 \delta R_\tau^{kl}  \,\equiv\,
 {R_{\tau,V+A}^{kl}\over |V_{ud}|^2} - {R_{\tau,S}^{kl}\over |V_{us}|^2}\, .
\end{equation}
%
The only non-zero contributions are proportional 
to the mass-squared difference $m_s^2-m_d^2$ or to vacuum expectation
values of SU(3)-breaking operators such as $\delta O_4
\equiv \langle 0|m_s\bar s s - m_d\bar d d|0\rangle \approx (-1.4\pm 0.4)
\cdot 10^{-3}\; \mathrm{GeV}^4$ \cite{PP:99,GJPPS:05}. The dimensions of these operators
are compensated by corresponding powers of $m_\tau^2$, which implies a strong
suppression of $\delta R_\tau^{kl}$ \cite{PP:99}:
\beqn\label{eq:dRtau}
 \delta R_\tau^{kl} &\!\!\approx &\!\!  24\, S_{\mathrm{EW}}\; \left\{ {m_s^2(m_\tau^2)\over m_\tau^2} \,
 \left( 1-\epsilon_d^2\right)\,\Delta_{kl}(\alpha_s)
 \right.\no\\ &&\hskip 1.3cm\left.
 - 2\pi^2\, {\delta O_4\over m_\tau^4} \, Q_{kl}(\alpha_s)\right\}\, ,
\eeqn
where $\epsilon_d\equiv m_d/m_s = 0.053\pm 0.002$ \cite{LE:96}.
The perturbative 
corrections $\Delta_{kl}(\alpha_s)$ and
$Q_{kl}(\alpha_s)$ are known to $O(\alpha_s^3)$ and $O(\alpha_s^2)$,
respectively \cite{PP:99,BChK:05}.

The $J=0$ contribution to $\Delta_{00}(\alpha_s)$ shows a rather
pathological behaviour, with clear signs of being a non-convergent perturbative
series. Fortunately, the corresponding longitudinal contribution to
$\delta R_\tau\equiv\delta R_\tau^{00}$ can be estimated phenomenologically with a much better
accuracy, $\delta R_\tau|^{L}\, =\, 0.1544\pm 0.0037$ \cite{GJPPS:05,JOP:06},
because it is dominated by far by the well-known $\tau\to\nu_\tau\pi$
and $\tau\to\nu_\tau K$ contributions. To estimate the remaining transverse
component, one needs an input value for the strange quark mass. Taking the
range
$m_s(m_\tau) = (100\pm 10)\:\mathrm{MeV}$ \
[$m_s(2\:\mathrm{GeV}) = (96\pm 10)\:\mathrm{MeV}$],
which includes the most recent determinations of $m_s$ from QCD sum rules
and lattice QCD \cite{JOP:06},
one gets finally $\delta R_{\tau,th} = 0.216\pm 0.016$ \cite{PI:07b}, which implies
\beqn\label{eq:Vus_det}
 |V_{us}| &=& \left(\frac{R_{\tau,S}}{\frac{R_{\tau,V+A}}{|V_{ud}|^2}-\delta
 R_{\tau,\mathrm{th}}}\right)^{1/2}
 \no\\ &=&
  0.2166\pm 0.0019_{\mathrm{\, exp}}\pm 0.0005_{\mathrm{\, th}}\, .
\eeqn
A larger central value,
$|V_{us}| = 0.2217\pm 0.0032$,
is obtained with the old world average for $R_{\tau,S}$.

Sizeable changes on the experimental determination of $R_{\tau,S}$ could be expected from
the full analysis of  the huge BaBar and Belle data samples. In particular, the high-multiplicity
decay modes are not well known at present.
The recent decrease of several experimental tau branching ratios is also worrisome.
As pointed out by the PDG \cite{PDG}, 15 of the 16 branching fractions measured at the B factories  are smaller than the previous non-B-factory values. The average normalized difference between the two sets of measurements is $-1.36\,\sigma$.
Thus, the result (\ref{eq:Vus_det}) could easily fluctuate in the near future.
In fact, combining the measured Cabibbo-suppressed $\tau$ distribution with electroproduction data,
a slightly larger value of  $|V_{us}|$ is obtained \cite{MA:10}.

The final error of the $V_{us}$ determination from
$\tau$ decay is dominated by the experimental uncertainties. If $R_{\tau,S}$
is measured with a 1\% precision, the resulting $V_{us}$ uncertainty will
get reduced to around 0.6\%, i.e. $\pm 0.0013$, making $\tau$ decay the best source of
information about $V_{us}$.

An accurate measurement of the invariant-mass distribution of the final hadrons
could make possible a simultaneous determination
of $V_{us}$ and the strange quark mass, through a correlated analysis of
several weighted differences $\delta R_\tau^{kl}$. However, the extraction of $m_s$ suffers from
theoretical uncertainties related to the convergence of the perturbative series
$\Delta_{kl}(\alpha_s)$. A better
understanding of these corrections is needed.

\section{CHIRAL SUM RULES}

When $m_{u,d,s}=0$, the QCD Lagrangian has an independent
$SU(3)$ flavour invariance for the left and right quark quiralities. The two quiralities
have exactly the same strong interaction, but they are completely decoupled.
This chiral invariance guarantees that the two-point correlation function of a left-handed and a right-handed quark currents, $\Pi_{LR}(s) = \Pi^{(0+1)}_{ud,V}(s) - \Pi^{(0+1)}_{ud,A}(s)$,
vanishes identically to all orders in perturbation theory (the vector and axial-vector
correlators receive identical perturbative contributions).
The non-zero value of $\Pi_{LR}(s)$ originates in the spontaneous breaking of chiral symmetry by the QCD vacuum. At large momenta, the corresponding OPE only receives contributions from operators with dimension $d\ge 6$,
%
\bel{eq:PiLR}
\Pi_{LR}^{\mathrm{OPE}}(s) = -\frac{\cO_6}{s^3}+\frac{\cO_8}{s^4}+\cdots
\ee
The non-zero up and down quark masses induce tiny
corrections with dimensions two and four, which are negligible at high energies.

At very low momenta,
Chiral Perturbation Theory ($\chi$PT) dictates the low-energy expansion of $\Pi_{LR}(s)$
in terms of the pion decay constant and the $\chi$PT couplings $L_{10}$ [$\cO(p^4)$] and $C_{87}$
[$\cO(p^6)$].

Analyticity relates the short- and long-distance regimes through the dispersion relation
\beqn\label{eq:dispersion}
\lefteqn{\frac{1}{2\pi i} \oint_{|s|=s_0} \! ds\, w(s)\, \Pi_{LR}(s)\: =\,
 -\int_{s_{th}}^{s_0} ds\, w(s)\, \rho(s)} &&
\no\\ &&\mbox{}\!\!
 + 2 f_\pi^2\, w(m_\pi^2) + \mathrm{Res}\!\left[w(s)\Pi_{LR}(s),s=0\right]\, ,
\eeqn
where $\rho(s)\equiv\frac{1}{\pi}\,\mathrm{Im}\Pi_{LR}(s)$ and $w(s)$ is an arbitrary weight function that is analytic in the whole complex plane except in the origin (where it can have poles). The last term in
(\ref{eq:dispersion}) accounts for the possible residue at the origin.

For $s_0\le m_\tau^2$, the integral along the real axis can be evaluated
with the measured tau spectral functions.
Taking $w(s) = s^n$ with $n\ge 0$, there is no residue at the origin and,
with $s_0$ large enough so that the OPE can be applied in the entire circle $|s|=s_0$,
the OPE coefficients are directly related to the spectral function integration. With $n=0$ and 1,
there is no OPE contribution in the chiral limit and one gets the celebrated first and second
Weinberg sum rules \cite{Weinberg:1967kj}.
For negative values of the integer $n$, the OPE does not contribute either while the residues at zero
are determined by the $\chi$PT low-energy couplings, which can be then experimentally determined
\cite{DGHS:98}.

Moreover, the absence of perturbative contributions makes (\ref{eq:dispersion}) and ideal tool to investigate possible quark-hadron duality effects, formally defined through
\cite{Shifman:2000jv,Cata:2005zj,GON07}
\beqn
\label{eq:DV}
\lefteqn{{\rm DV}_w\,
 \equiv \:\frac{1}{2 \pi i} \, \oint_{|s|=s_0} ds\; w(s) \left( \Pi_{LR}(s) - \Pi^{\rm{OPE}}_{LR}(s)\right)}&&
\no\\
&& \;\; =\; \int^{\infty}_{s_0} \mathrm{d}s~w(s)~\rho(s)\, . \hskip 3.4cm\mbox{}
\eeqn
This has been thoroughly studied in Ref.~\cite{GonzalezAlonso:2010rn},
using for the spectral function beyond $s_z \sim 2.1~\mathrm{GeV}$
the parametrization \cite{CGP:09,Shifman:2000jv,Cata:2005zj,GON07}
\be\label{eq:model}
\rho(s\ge s_z) \, =\, \kappa~ e^{-\gamma s} \sin(\beta (s-s_z))\, ,
\ee
and finding the region in the 4-dimensional $(\kappa,\gamma,\beta,s_z)$ parameter space
that is compatible with the most recent experimental data \cite{ALEPH:05} and the following theoretical constraints at $s_0\to\infty$: first and second Weinberg sum sules \cite{Weinberg:1967kj} and the sum rule of Das et al. \cite{Das:1967it} that determines the pion electromagnetic mass difference.

Ref.~\cite{GonzalezAlonso:2010rn} performs a statistical analysis, scanning the parameter space $(\kappa,\gamma,\beta,s_z)$ and selecting those `acceptable' spectral functions which satisfy the experimental and theoretical constraints. From a generated initial sample of $160,000$ tuples, one
finds $1,789$ acceptable distributions compatible with QCD and the data.
The differences among them determine how much freedom is left for the behaviour of the spectral function beyond the kinematical end of the $\tau$ data. For each acceptable spectral function one calculates
the parameters $L_{10}$, $C_{87}$, $\cO_6$ and $\cO_8$, obtained through
the dispersion relation (\ref{eq:dispersion}) with the appropriate weight functions. The resulting
statistical distributions determine their finally estimated values; the dispersion of the numerical results provides a good quantitative assessment of the actual uncertainties.

The study has been also performed with pinched weight functions of the form $w(s) = s^n (s-s_z)^m$ ($m>0$) that vanish at $s=s_z$. As expected, these weights are found to minimize the uncertainties from duality-violation effects, allowing for a more precise determination of the hadronic parameters.
One finally obtains \cite{GonzalezAlonso:2010rn},
\beqn\label{eq:hadrResults}
&& \hskip -.7cm L_{10}^r(M_\rho) \, =\, -(4.06\pm 0.39)\cdot 10^{-3}\, ,
\no\\
&& \hskip -.7cm C_{87}^r(M_\rho) \, =\, (4.89\pm 0.19)\cdot 10^{-3}\;\mathrm{GeV}^{-2}\, ,
\no\\
&& \hskip -.7cm \cO_6 \, =\, (-4.3\,{}^{+\, 0.9}_{-\, 0.7})\cdot 10^{-3}\;\mathrm{GeV}^{6}\, ,
\no\\
&& \hskip -.7cm \cO_8 \, =\, (-7.2\,{}^{+\, 4.2}_{-\, 5.3})\cdot 10^{-3}\;\mathrm{GeV}^{8}\, .
\eeqn
%

\begin{figure}[t]
\centerline{
\begin{minipage}[t]{.3\linewidth}\centering
\centerline{\includegraphics[width=7.cm]{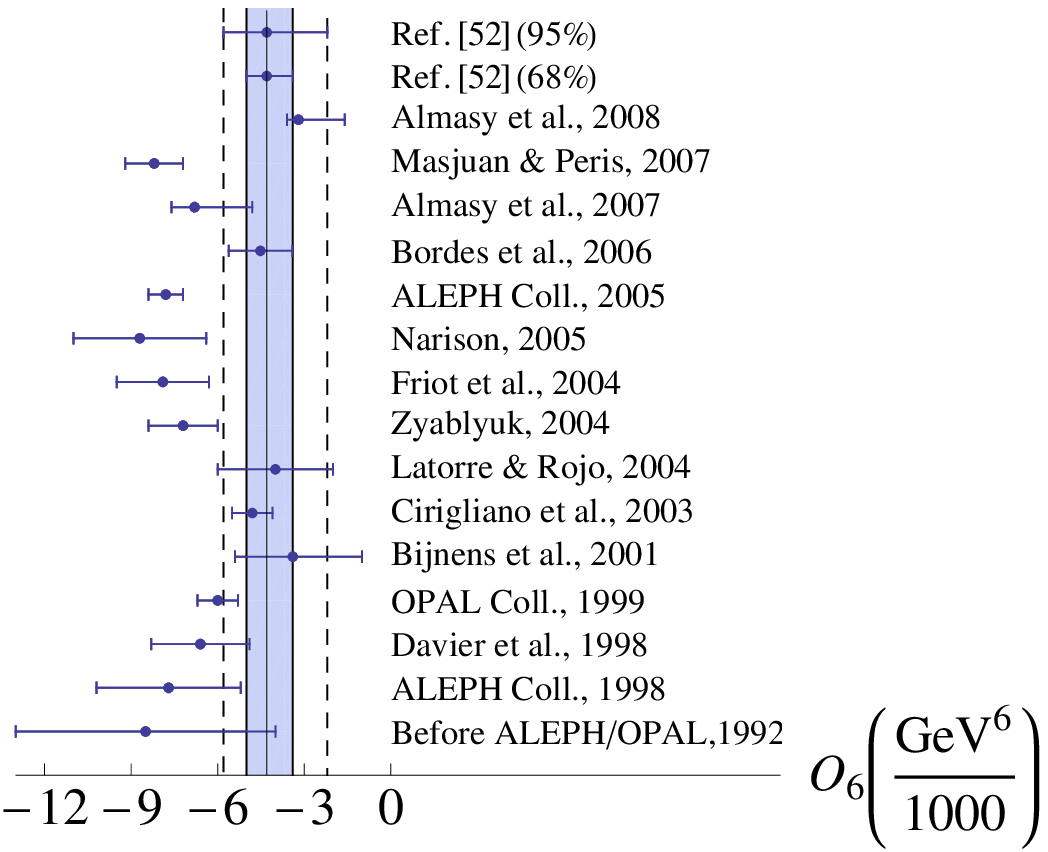}}
\end{minipage}
}
\vspace{0.5cm}
\centerline{
\begin{minipage}[t]{.3\linewidth}\centering
\centerline{\includegraphics[width=7.cm]{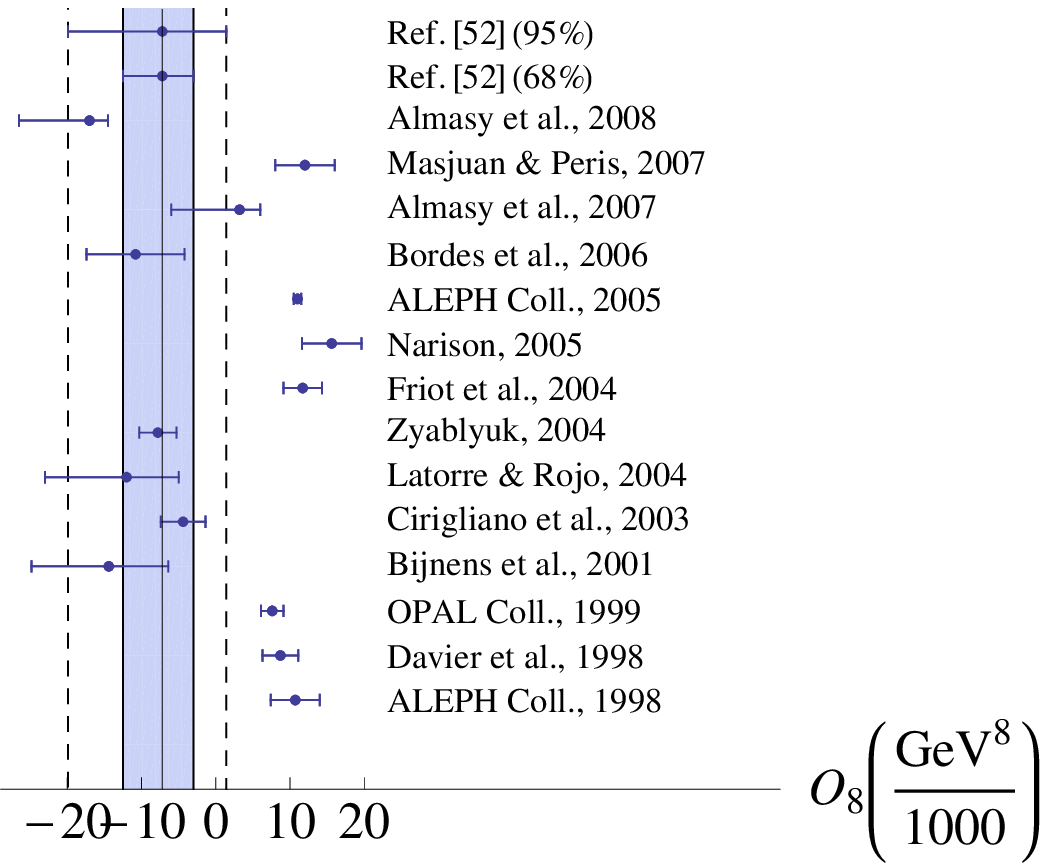}}
\end{minipage}
}
\vskip -.5cm
\caption{Published results for $\cO_6$ 
and $\cO_8$ \cite{GonzalezAlonso:2010rn}. 
}
\label{fig:comparisonPW2}
\end{figure}

The determination of the two $\chi$PT couplings is in good agreement with (but more precise than)
recent theoretical
calculations, using Resonance Chiral Effective Theory \cite{RChT} and large--$N_C$ techniques at the next-to-leading order \cite{PI:02}, which predict \cite{PRS:08}:
$L_{10}^r(M_\rho)  = -(4.4\pm 0.9)\cdot 10^{-3}$ and
$C_{87}^r(M_\rho)  = (3.6\pm 1.3)\cdot 10^{-3}\;\mathrm{GeV}^{-2}$. It also agrees with the present
lattice estimates of $L_{10}^r(M_\rho)$ \cite{Shintani:2008qe}.

Duality-violation effects have very little impact on the determination of $L_{10}$ and $C_{87}$ because
the corresponding sum rules are dominated by the low-energy region where the data sits. Thus, one obtains basically the same results with pinched and non-pinched weight functions. This is no-longer true for $\cO_6$ and $\cO_8$, which are sensitive to the high-energy behaviour of the spectral function; pinched-weights provide then a much better accuracy. This could explain the numerical differences among previous estimates
\cite{ALEPH:05,OPAL:98,DGHS:98,Bijnens:2001ps}, shown in Fig.~\ref{fig:comparisonPW2},
where duality violation uncertainties were not properly assessed. The
results (\ref{eq:hadrResults}) fix with accuracy the value of $\cO_6$ and determine
the sign of $\cO_8$. This information is needed to calculate the electromagnetic penguin contribution
to the CP-violating ratio $\varepsilon'_K/\varepsilon_K$ 
\cite{GPP:11}.


\section*{ACKNOWLEDGEMENTS}
I would like to dedicate this work to the memory of our friend and collaborator Ximo Prades,
who sadly passed away recently. Ximo has made many relevant contributions to the physics of the tau lepton,
some of which have been mentioned here.
This work has been supported
by MICINN, Spain (grants FPA2007-60323 and
Consolider-Ingenio 2010 CSD2007-00042, CPAN) by the
EU Contract MRTN-CT-2006-035482 (FLAVIAnet) and by Generalitat Valenciana
(PROMETEO/2008/069).


\end{document}